\documentclass[11pt]{article}

\usepackage[final]{acl}

\usepackage{times}
\usepackage{latexsym}
\usepackage[T1]{fontenc}
\usepackage{todonotes}
\usepackage[utf8]{inputenc}

\usepackage{microtype}
\usepackage{todonotes}
\usepackage{inconsolata}

\usepackage{graphicx}
\usepackage{graphicx}

\usepackage{subcaption}
\usepackage{longtable}
\usepackage{booktabs}
\usepackage{multirow}  
\usepackage{hyperref}
\usepackage[most]{tcolorbox}
\usepackage{multicol}
\usepackage{pdfpages}
\usepackage{array}

\usepackage{todonotes}
\title{Towards a Benchmark for Fire Safety Code Interpretation \\ in the Hong Kong AEC Industry}

\author{
 \textbf{Le Qiu\textsuperscript{1}},
 \textbf{Zhuan Anne Zhang\textsuperscript{2, 3}},
 \textbf{Hok Ming Adeline Chan\textsuperscript{2, 3}},
 \textbf{Emmanuele Chersoni\textsuperscript{1, 4}}
\\
 \textsuperscript{1}Language Science and Technology, Hong Hong Polytechnic University, Hong Kong \\
 \textsuperscript{2}AAL INNOVATION Limited, Hong Kong\\
 \textsuperscript{3}Journalism and Communication, Hong Kong Shue Yan University, Hong Kong\\
 \textsuperscript{4}Artificial Intelligence and the Humanities, Hong Hong Polytechnic University, Hong Kong
\\
 \texttt{lani.qiu@connect.polyu.hk, anne@aalinnovationai.com}
 \\
 \texttt{adeline@aalinnovationai.com, emmanuele.chersoni@polyu.edu.hk}
}

\begin{document}
\maketitle
\begin{abstract}
Large Language Models (LLMs) are increasingly explored in the AEC (Architecture, Engineering, and Construction) domain, yet their reliability in regulatory tasks remains uncertain.% 

In this study, we focused on \textit{Hong Kong fire safety codes}: together with domain experts, we create a localized dataset of multiple‑choice (MCQs) and open‑ended questions (OEQs) to evaluate 12 widely used LLMs, including both proprietary and open‑source models.

Our findings show that proprietary models may achieve 80–90\% accuracy on MCQs, while open‑source models generally lag behind. On OEQ, humans rated the answers at 50-70\% accuracy, suggesting that LLMs in specialized compliance tasks still have large margins of improvement. %We also observed that enabling \textit{thinking} did not significantly improve performance. 

%This work contributes a benchmark for fire safety code analysis and we hope the benchmark and findings can inspire similar studies across other AEC subdomains, ultimately advancing the safe and practical deployment of LLMs in AEC applications.
\end{abstract}

\section{Introduction}

The AEC industry is one of the many domains in which Generative AI and LLMs are rapidly gaining momentum: the most tech-savvy architects have started to adopt LLM-based tools in their everyday design practice, while clients increasingly see the usage of AI as a guarantee for their return of investment \cite{del2022machine}, as well as a valuable tool for workflows such as cost prediction and quality control \cite{kampelopoulos2025review}.
 An open question is whether current state-of-the-art LLMs can already be reliable support tools for specialized and safety-critical AEC contexts.

%\par 
In Tai Po, in the New Territories of the Hong Kong Special Administrative Region, a major and tragic fire incident has recently brought renewed public attention to the importance of fire safety in urban environments\footnote{\url{https://en.wikipedia.org/wiki/Wang_Fuk_Court_fire}}. For us, as NLP practitioners, it should be a natural question to ask, whether modern technologies can help us to prevent such type of incident from occurring again in the future.

Motivated by this context, we take fire safety as a critical lens to evaluate the current capabilities of state-of-the-art AI models within the AEC sector. Although prior research has examined LLMs in AEC‑related tasks --- such as code interpretation and design assistance \cite[e.g.][]{guo2025toward,liang2026aecbench} --- systematic evaluations specifically targeting fire safety remain scarce. %\todo[inline]{EMM: to be checked in recent literature}
%%% why fs 
Fire safety represents a high‑stakes AEC subdomain that relies heavily on precise interpretation of regulatory codes, making it an ideal testbed for assessing whether AI can reliably support professional decision‑making in AEC. Building upon this motivation, we have collaborated with domain experts and constructed a Q\&A (Question Answering) dataset relevant to local fire safety practice in Hong Kong. Using this dataset, we have evaluated both proprietary LLMs and open‑source models to understand their strengths, limitations, and failure modes in this high‑stakes domain. 
\par In summary, our work primarily make two contributions:
i)  We present the first dataset grounded in the Hong Kong fire safety regulatory context, offering a structured benchmark for evaluating AI systems in one of the most safety‑critical AEC subdomains; 
ii) we provide a systematic evaluation of LLMs on code compliance checking tasks in Hong Kong fire safety, analyzing their capabilities, reliability gaps and resource consumptions when interpreting real‑world regulatory codes.\footnote{Our dataset and code are available at \url{https://github.com/Laniqiu/HK-FSCode-LLM-Eval}.}
%Our findings serve as reference for academic research and practical applications in relevant domains, and offer insights into the feasibility of developing future vertical models for AEC. %, showing how smaller open‑source models perform relative to large proprietary systems and identifying the domain‑specific challenges that specialized models must address.
    % \item We offer insights into the feasibility of developing future vertical models for AEC, showing how smaller open‑source models perform relative to large proprietary systems and identifying the domain‑specific challenges that specialized models must address.%\todo[inline]{anne:I feel this item also make sense. Shall we say this evaluation could provide reference baseline for LLM fine-tuning?}
    % \todo[inline]{merge with 2?}
% \end{enumerate}

% The following sections describe related work, dataset creation, evaluuatithe model selection, prompt design, and evaluation metrics used in our experiments.

\section{Related Work} 
% llm in aec?
\subsection{LLMs in the AEC Industry}

AEC workflows are inherently document-intensive, relying on large volumes of unstructured  data such as design briefs and safety regulations. This makes the industry particularly receptive to automation technologies.  Evidence shows that recent LLMs have been swiftly integrated into various AEC tasks,  ranging from basic document processing --- such as information extraction and summarization --- to more advanced applications like regulatory compliance inspection \cite{kampelopoulos2025review}. Commercial general‑purpose LLMs—such as ChatGPT, Gemini, and Claude --- are among the most widely adopted, due to their strong reasoning capabilities and ready-to-use features.

However, a more widespread application of LLMs in the AEC industry faces several obstacles. The most prominent issue is domain knowledge: AEC tasks often require extensive expertise spanning multiple disciplines, including architecture, structural engineering, mechanical and electrical engineering (MEP), and project management \citep{buildings14010220}, and even for general‑purpose LLMs it is difficult to reliably cover such specialized and multidisciplinary knowledge. Another widely discussed challenge are hallucinations \citep{huang2023survey}, where AI systems generate plausible‑sounding but incorrect information. In safety‑critical AEC subdomains—such as fire safety engineering, structural safety assessment, or code compliance checking—hallucinations pose significant risks \citep{xiong2026socio}, as incorrect interpretations may lead to unsafe decisions.

\subsection{LLMs for Compliance Checking}

In response to these limitations, researchers have developed domain-specific LLMs tailored for AEC, mainly for compliance checking \citep{zheng2023llm,zheng2025automating}. For example, CivilGPT and SiKong \footnote{CivilGPT: \url{https://civilgpt.tongji.edu.cn/};\\ SiKong: \url{https://github.com/SikongSphere/sikong}} are two models developed for the Chinese AEC. CivilGPT specializes in civil engineering knowledge, while SiKong is a fine‑tuned model based on LLaMA and Alpaca, designed to support architecture‑related question answering tasks. Both focus on the knowledge systems specific to Mainland China. 
%\todo{TO BE INTEGRATED HERE, Other possibly similar dataset on fire safety codes: Accord-NLP and SearchFireSafety }
%\todo[inline]{EMM: maybe to be integrated with the literature on LLMs for AEC and knowledge management?}

Notably,  AEC regulations exhibit strong regional specificity, as codes and regulations vary across jurisdictions: for example, a system like SiKong may perform well on the Chinese regulatory frameworks, but its applicability is limited when dealing with the codes of other regions. While this issue implies the need for localization in AEC-oriented LLMs, AEC‑specific datasets remain scarce and qualified human experts are extremely difficult to recruit for dataset construction. 

Compliance checking on fire safety codes in NLP is not a common research topic, although some benchmarks exist, such as the CODE-ACCORD corpus, containing regulatory codes from the United Kingdom and Finland and also including fire safety ones \citep{hettiarachchi-etal-2026-snowtec}; and SearchFireSafety, a dataset for statute-centric legal question-answering centered on South Korean fire-safety regulations \citep{chae2026evaluating}. 

To our knowledge, no specific NLP research has been conducted dedicated to Hong Kong fire safety codes. Since developing a dataset of instructions to fine-tune a model specifically for the Hong Kong fire safety regulation is not practical due to the costs and the limited applicability, we focus our efforts on constructing an evaluation-only dataset that can be used to test the performance of state-of-the-art LLMs, to see if their exposure to massive scales of data and advanced reasoning capabilities can compensate for the lack of localization.

\section{Dataset Creation}

We constructed a Q\&A dataset sourced from \textit{Code of Practice for Fire Safety in Buildings 2011} (FS Code) \footnote{While the FS Code is available in both English and Cantonese, we used the 2024 English version to align with local qualification examinations, which are conducted in English. Source: \url{https://www.bd.gov.hk/doc/en/resources/codes-and-references/code-and-design-manuals/fs_code2011.pdf}}, published by the Hong Kong Buildings Department. The FS Code is the central regulatory document governing fire safety provisions in Hong Kong and serves as the official reference for both professional practice and local qualification examinations, making it a reliable foundation for assessing LLM understanding of domain‑specific fire safety knowledge. Our samples comprehensively cover both core components of the code: an \textbf{architectural} component that includes means of escape, means of access, and passive fire protection measures (e.g., fire-resisting construction, fire barriers, etc.), and a \textbf{fire engineering} component that specifies the performance and design of active fire protection systems and fire service installations (e.g., automatic sprinklers, fire detection and alarm systems, etc.). Following the style of the existing HKIA/ ARB Professional Assessment (PA)\footnote{The PA is the official professional qualification examination administered by The Hong Kong Institute of Architects (HKIA) and the Architects Registration Board (ARB). It is the primary pathway to obtain professional membership in HKIA and to become registered architects in Hong Kong.}, we designed two major categories of questions: multiple‑choice questions and open‑ended questions. %Both question types are organized on multiple levels, depending on the type of operation that a model is supposed to perform to answer, e.g. from factual recall to scenario-based applications that may require several reasoning steps (see Appendix, Table \ref{tab:dif_levels}). %The MCQs span 3 difficulty levels, ranging from factual recall to scenario‑based application, and the OEQs 4 levels, from short factual responses to complex multi‑step reasoning and scenario analysis (see Appendix, Table \ref{tab:dif_levels}). 
%\textbf{EMM: can we be sure about the difference in difficulty? Maybe they just require different skills... I was wondering, was performance high even when the prompt did not specify HK as a context?}

We employed a hybrid AI-human workflow for Q\&A generation. We first used the recent, state-of-the-art GPT‑5 \cite{singh2025openai} model to generate candidate questions --- either by prompting the model with specific requirements or by supplying relevant excerpts from the FS Code as source material. Near‑identical or overlapping questions were manually removed. Following this stage, each remaining question underwent comprehensive human verification, including review and correction of the question, answer and references. %\todo{EMM: how many humans checked each question? Any policy to solve disagreements? We should add this info here ->  lani: the qa pairs were first roughly checked by anne, then dennis, an expert, and then anne would check again, no specific resolution for disagreement was involved, i am afraid} 
More specifically, the expert review was conducted by two architects and one fire engineer, all of whom are Hong Kong‑registered, experienced, and actively practicing professionals, with responsibilities divided according to domain expertise: specialists in fire safety regulations examined the architectural components, while professionals with relevant technical backgrounds reviewed the engineering components. Each question would be reviewed by two experts and retained only if both consider it valid; items on which there was disagreement were discarded from the dataset. %Each question would receive at least two reviews. 
%\todo{EMM: if we have one check of the expert and one by Anne, can we compute agreement between Anne and the expert at least? -> it's unlikely, as they didn't leave specific scores/ ratings for calculation, but the disagreed items would generally be removed}
%about 2/3 are architectural and about 1/3 are engineering-related. Among the OEQs, 88 samples are architectural and 12 are engineering-related. 

The final version contains a total of 500 Q\&A pairs,  comprising 400 MCQs and 100 OEQs. Examples are provided in Appendix Table \ref{tab:data_exp}. %The MCQs cover 3 difficulty levels with sample counts of 131, 123, and 146, while the OEQs span 4 levels with counts of 15, 35, 40, and 10, respectively. In the MCQs, about one third of the items are architectural and two thirds are engineering-related, while in the  OEQs, this ratio is about 7:1. Examples are provided in Table \ref{tab:data_exp}. \todo[inline]{EMM: OEQ are very imbalanced, maybe we can just regroup them putting together 1 and 2 vs. 3 and 4  -> lani: how about we don't mention about diff levels? instead, EMM: ok, so we just present the results without separation between diff levels? I am ok with it, it will solve a lot of issues}

\section{Evaluation Settings}

\subsection{Models}

We evaluated 12 distinct LLMs, as outlined in Table \ref{tab:model_choice}: proprietary models (top) include Claude Opus \cite{anthropic2024opus}, DeepSeek \cite{deepseek2024}, Gemini 3  \cite{gemini32025}, GPT‑4  \cite{hurst2024gpt}, Grok‑4 \cite{xai2024grok4} and Qwen‑Plus \cite{qwen36plus}; open-sourced models (bottom) include Qwen2.5-7B \cite{qwen2.5, qwen2}, Llama3-8B, Mistral-7B, DeepSeek-7B, Llama3-70B and Qwen3-80B \cite{qwen3technicalreport, qwen2.5-1m}. For DeepSeek and Gemini 3, both \textit{thinking} and \textit{non-thinking} modes are evaluated and each is counted as a separate evaluation, bringing the total to 14.

The selection was based on several considerations. First, it is impractical to include a large amount of available LLMs, given the time and financial constraints. Still, we aimed to cover the popular and representative LLMs that the AEC practitioners are likely to use in real‑world applications. Second, we intentionally included smaller‑scale models to examine whether model size affects performance. Third, we included both \textit{thinking} and \textit{non-thinking} models for DeepSeek and Gemini 3, trying to investigate whether explicit reasoning‑mode capabilities lead to measurable differences. 

Proprietary models are accessed via APIs and open-source models are run through Hugging Face Transformers. We use default inference parameters for all models, except that we set the maximum number of newly generated tokens to 512 to ensure comparable answer lengths.\footnote{This constraint only applies to the final answer returned
to the user, as the internal reasoning tokens in the thinking
mode are not counted in the generation limit. Note that the
specific parameter names may vary across providers.}

\begin{table*}[tb!]
    \centering
    \caption{A summary of model identifiers and pricing\protect\footnotemark} 
    \label{tab:model_choice}
    \begin{tabular}{l|l|l}
    \toprule
     & Identifier & In/ Out Price \\\midrule
      Claude Opus & claude-4.5-opus & 5/ 25\\%claude-opus-4-5-20251101 \\
      % DeepSeek (Thinking) & deepseek-v3.2 (thinking)\\
      % DeepSeek (Non-thinking) & deepseek-v3.2 (non-thinking)\\
      % DeepSeek &  deepseek-v3.2 (thinking + non-thinking)\\
      DeepSeek & deepseek-chat & 0.007$\sim$ 1.32 / 0.66$\sim$ 3.96  \\ % deepseek-v3.2
    %   Gemini 3 (Thinking) & gemini-3-flash-preview (thinking)\\
    % Gemini 3 (Non-thinking) & gemini-3-flash-preview (non-thinking)\footnotemark\\
    Gemini 3 & gemini-3-flash-preview & 0.05 $\sim$ 1 / 3\\
      GPT-4 & gpt-4.1 & 0.5 $\sim$ 2 / 8 \\ %2025-04-14 \\
      Grok-4 & grok-beta & 0.2 / 0.5\\
      Qwen Plus  & qwen-plus & 0.4 $\sim$ 2 / 1.6 $\sim$ 6\\
      \midrule
      Qwen2.5-7B & Qwen/Qwen2.5-7B-Instruct & N.A. \\
      Llama3-8B & meta-llama/Llama-3.1-8B-Instruct & N.A.\\
      Mistral-7B  & mistralai/Mistral-7B-Instruct-v0.3 & N.A.\\
    DeepSeek-7B & deepseek-ai/deepseek-llm-7b-chat & N.A. \\
     Llama3-70B  & meta-llama/llama-3.3-70b-instruct & N.A.\\
   Qwen3-80B &  Qwen/Qwen3-Next-80B-A3B-Instruct & N.A.\\       
    %  \midruleGPT-5 \cite{singh2025openai} & gpt-5 \\%-2025-08-17 \\
      \bottomrule
    \end{tabular}
   % \label{models}
\end{table*}

\footnotetext{Prices are quoted per 1 million tokens in USD and provided for reference only (actual costs may vary depending on usage time, mode, cache status and membership tier); The thinking and non‑thinking modes for DeepSeek and Gemini 3 are implemented within the same model architecture, with the mode selection controlled by API parameters.}%For Gemini 3, the thinking version corresponds to a default \textit{thinking\_level = high}, while the non‑thinking version is obtained by explicitly setting \textit{thinking\_level = minimal}. Documentation: \protect\url{https://api-docs.deepseek.com/guides/thinking_mode} and \protect\url{https://ai.google.dev/gemini-api/docs/gemini-3}.}

\begin{table*}[htb!]
    \centering
    \small
    \caption{Human evaluation rubric: dimensions, descriptors and scoring ranges.}
    \label{tab:human_eval_rubrics}
    \begin{tabular}{p{0.2\linewidth}|p{0.7\linewidth}}
    \toprule
       \multicolumn{2}{c}{Accuracy}\\
    \midrule
    Fully accurate & The response is correct, complete, and aligned with the Hong Kong Fire Safety Code (100\%). \\ \hline
     Minor inaccuracy & The response is generally correct but contains small inaccuracies or omissions that do not fundamentally affect compliance (90\%).\\ \hline
     Major issues & The response contains substantial problems. This level is further divided into four sub‑categories: \\ \hline
            & Irrelevant but real clause cited (80\%). \\ 
            &  Omission of critical information (60\%).\\ 
            &  Hallucination or fabricated content (30\%)\\ 
            &  Fundamental misunderstanding of the regulation (0\%).\\
    \midrule
      \multicolumn{2}{c}{Clarity}\\
      \midrule
Clear and professional &  Fluent, logical language with precise domain terminology (100\%). \\\hline 
Minor clarity issues &  Slightly awkward phrasing or redundancy, but still comprehensible  (90\%). \\\hline  
 Inaccurate or inconsistent terminology & Uses incorrect, imprecise, or non-standard Hong Kong fire safety terms (80\%). \\\hline 
 Logical gaps or weak reasoning & Disconnected arguments, logical jumps, or unsupported conclusions (50\%). \\\hline 
 Insufficient explanation / justification & Conclusions lack specific clause references, rationale, or step-by-step reasoning (50\%). \\\hline 
 Poor readability / disorganized structure & Difficult to follow due to lack of structure, paragraphing, or logical flow  (0\%). \\
     \bottomrule
    \end{tabular}
\end{table*}

\subsection{Metrics}

For MCQs, we use \textbf{accuracy} as a standard metric, i.e. the proportion of questions for which the
model selects the correct option. For OEQs, which involve free-form text, we use both automatic and
human evaluation.

Generally, automatic evaluation operates by quantifying the textual similarity between the model-generated output and the reference answer. We employ standard metrics widely used in natural language generation tasks, including
BLEU \cite{papineni-etal-2002-bleu}, METEOR \cite{banerjee-lavie-2005-meteor}, ROUGE-L \cite{lin-2004-rouge} and BERTScore \cite{zhang2020bertscore}. These metrics capture different aspects of similarity:
BLEU emphasizes n-gram precision, METEOR incorporates synonym matching, ROUGE-L focuses on longest-common subsequence recall, and BERTScore measures semantic similarity using contextual embeddings.%In the presented evaluation, BERTScore is computed using RoBERTa-large and reported with the F1 score, whereas BLEU is reported with n=4.
\footnote{Scores are computed using \textit{pycocoevalcap} for BLEU (n=4), METEOR and ROUGE; and \textit{bert\_score} for BERTScore (F1 scores from RoBERTa-large are reported) .}%\footnote{\url{https://huggingface.co/FacebookAI/
% roberta-large}} 

For human evaluation, we designed two rating dimensions that potentially mirror criteria commonly used in professional examinations: \textbf{accuracy} and \textbf{clarity}. Accuracy assess whether the 
response appropriately addresses the question and 
adheres to the facts; clarity measures the linguistic 
quality of the response — whether it is presented 
in a clear, fluent, coherent and professional manner. Each dimension was defined through multiple 
descriptors that specify different aspects of performance, as shown in Table \ref{tab:human_eval_rubrics}. Each descriptor 
carried a fixed penalty level. The selected descriptor determined the final score for that dimension. 
For example, a response classified as fully accurate 
would receive the full score (100\%) for accuracy, 
whereas a response categorized as having major 
issues might incur the maximum penalty and end 
up with a score of zero (0\%). 

Due to time constraints and the high cost associated with human assessment, we did not evaluate OEQ outputs for all models. Instead, we selected a subset of five models that performed strongly in the automatic evaluation stage and conducted human evaluation on their OEQ outputs. By this design, for each question, we have 5 Q\&A pairs (one for each of the 5 evaluated models), and each Q\&A pair would be independently and randomly evaluated by at least two raters. Rather 
than assigning numerical scores, raters are invited to classify each model response into the descriptors that best characterized its accuracy and clarity, without knowing the model identity. The judgment should be based on their professional knowledge and the provided reference answers. All raters are AEC professionals registered and practiced in Hong Kong. Architectural-related questions are evaluated by architects, and engineering-related questions were rated by fire engineers.

% Additionally, we adopt an \textbf{LLM-as-a-judge} approach to assess model outputs, following recent practices in automatic evaluation \cite{gu2024survey}. In the LLM-as-a-judge process, an LLM is employed as a rater to evaluate model outputs along the two dimensions defined above: accuracy and clarity. The prompts in Table \ref{qt:prompt_llm_eval} were designed to mirror the instructions given to human raters.

\subsection{Prompts}
The evaluation prompts follow a fixed format as shown in Table \ref{tab:prompts_eval}. The prompts explicitly indicate that the applicable jurisdiction is Hong Kong (Table \ref{tab:tmp_oeq}). For MCQs, however, we prepare a second set of prompts that does not specify the jurisdiction (Table \ref{tab:tmp_mq2}). By this design, we intend to examine whether the absence of regional awareness --- likely in real expert-LLM interactions --- affects model performance.

\section{Results}
\label{sec:eval}

The scores are presented in Table \ref{tab:res}, along with the average token cost  (including both input and output) and the processing time (in seconds). 

For the human evaluation of OEQs, we reported their average ratings in Table \ref{tab:res_human}. As noted earlier, we evaluated the outputs from five models: four proprietary models and one open-source model — Gemini 3 (Thinking and Non-Thinking variants), GPT-4, Claude Opus, and Qwen2.5-7B. The selection was based on multiple considerations, including their automatic evaluation scores on both MCQs and OEQs, resource consumption, and availability in Hong Kong, among other practical factors. Qwen2.5 was included to provide a term of comparison of what can be obtained with a smaller-size, open model. 

% The LLM-as-a-judge evaluation (Table \ref{tab:res_llm}), as a supplement to human evaluation, assesses the full OEQ outputs of these five LLMs on accuracy and clarity. Grok-4, which showed a particularly strong performance in the MCQ, was adopted as the LLM judge for the outputs. \footnote{It should be noticed that Grok-4, when generating text, tends to produce less standardized formats, requiring more efforts to produce the final answer, and has higher token consumption, possibly due to its real-time search feature. Given that we chose to use it as a judge, it goes without saying that it had to be excluded by the LLM-as-a-judge study to avoid self-evaluation biases \citep{panickssery2024llm,xu2024pride}.} %%\footnote{Although Grok-4 demonstrated impressive performance
%on MCQs, especially with the w/ HK context condition, it was not included among the five evaluated LLMs due to concerns about stability and resource consumption. Compared with
%others, Grok-4 occasionally produced refusal responses and
%its outputs format was less standardized, requiring more efforts
%to extract the final answer. Besides, it exhibited high token
%consumption likely due to its real-time search feature.}

Overall, the proprietary models outperform smaller open-source models, which is expected given their substantial differences in parameter counts.\footnote{Most proprietary LLMs do not disclose their parameter size, but it is generally assumed that they are very large, often in the hundreds of billions or even trillions.} The bigger open-source models (i.e. Llama3-70B and Qwen3-80B) are the closest to the level of performance of proprietary models, although they generally still lag behind in all metrics. Interestingly, for most automatic generation metrics the larger open models do not improve over the scores of their smaller versions.

\begin{table*}[htb!]
    \centering
    \caption{A summary of evaluation results. The (*) marker indicates the model operates in a thinking mode, otherwise non-thinking mode by default, time cost is in seconds.  The top scores are \underline{underlined}.}
    \label{tab:res}
    \begin{subtable}{0.55\textwidth}
    \centering
    \begin{tabular}{l|c|c|c}
       \toprule
         & Accuracy & Time Cost & Token Cost \\
        \midrule
         Qwen Plus & 79.5  &  6 &  372 \\
         Grok-4 &     \underline{90}   & 12 & 2728 \\
         % GPT-5 & 85   & 16 & 1159 \\
         GPT-4 &  80.5   & 3 & 265 \\
         Gemini 3 &   88.5   & 2 & 257 \\
         Gemini 3\textsuperscript{*} & 89.25   & 17 & 2254 \\
         DeepSeek &  80.75   & 7 & 266 \\
         DeepSeek\textsuperscript{*}  &  83.5   & 68 & 1718 \\
         Claude Opus &   86   & 6 & 2562 \\
         \midrule
         Mistral-7B &  69.25   & N.A. & N.A.\\
         Qwen2.5-7B &  75.5   & N.A. & N.A. \\
         Llama3-8B &  77   & N.A. & N.A.\\
         DeepSeek-7B &  66   & N.A. & N.A. \\
         Llama3-70B & 79 & 6 & 274 \\ 
         Qwen3-80B & 79 & 1&292 \\
        \bottomrule
        \end{tabular}
        \caption{Accuracy for MCQs (w/ HK)}\label{tab:res_mcq1}
    \end{subtable}
    \hspace{0pt}
       \begin{subtable}{0.4\textwidth}
    \centering
    \begin{tabular}{c|c|c}
        \toprule
         Accuracy & Time Cost & Token Cost \\
        \midrule
            81.75 & 5 & 366\\
            83.75 &  15 & 2881 \\
            % 82 & 15 & 1057 \\
            78.75 & 2 & 2454 \\
            85.75 & 33 & 2221 \\
            \underline{86} & 24 & 3082\\
            80.75 & 15 & 334 \\
            81 & 76 & 1770 \\
            81.25 & 6 & 366 \\
                \midrule
            67.25 & N.A. & N.A. \\
            73 & N.A. & N.A. \\
            73 & N.A. & N.A. \\
            60.75 & N.A. & N.A. \\
            77.75 (78.73)\protect\footnotemark & 8 & N.A. \\%271 \\  
            78.25 & 2 & N.A.\\%290 \\
        \bottomrule
        \end{tabular}
        \caption{Accuracy for MCQs (w/o HK)}\label{tab:res_mcq2}
    \end{subtable}
    \begin{subtable}{\textwidth}
    \centering
        \begin{tabular}{l|c|c|c|c|c|c}
        \toprule
                    & BERTScore & BLEU & METEOR & ROUGE & Time Cost & Token Cost\\
          \midrule
           Qwen Plus & 84.6 & 3.42 & 16.23 & 15.35  &  6 & 356 \\
        Grok-4 & 84.41 & 3.19 & 19.78 & 15.97  &  16 & 3280 \\
        % GPT-5 & 84.26 & 2.24 & \underline{20.5} & 14.34  &  50 & 3322 \\
        GPT-4 & 85.8 & \underline{4.66} & 17.1 & \underline{17.74}  &  4 & 298 \\
        Gemini 3 & 85.41 & 3.43 &14.36 &16.8  &  2 & 281 \\
        Gemini 3\textsuperscript{*} & 85.55 & 4.03 & 15.16 & 17.53  &  15 & 1346 \\
        DeepSeek & 85.58 & 3.64 & 13.75 &  16.33  &  9 & 270 \\
        DeepSeek\textsuperscript{*}  & 85.56 & 4.34 &  15.6 & 16.4  &  56 & 1375 \\
        Claude Opus & 85 & 3.14 & \underline{19.25} & 15.86  &  10 & 2472 \\
        \midrule
        Mistral-7B &  85.2 &  3.97 &  15.08 & 16.15  &  N.A. & N.A. \\
        Qwen2.5-7B & \underline{85.63} & 3.21 & 12.93 & 16.52  &  N.A. & N.A. \\
        Llama3-8B & 84.9 & 3.68 &14.23& 16.69 &  N.A. & N.A. \\
        DeepSeek-7B & 84.91 & 2.76 & 10.89&  15.63  &  N.A. & N.A. \\ 
         Llama3-70B & 84.46 & 0.73 & 7.32 & 12.41 & 8 & N.A.\\%271 \\
         Qwen3-80B & 84.90 & 1.63 & 9.55 & 13.45 & 2 & N.A.\\%319 \\
            \bottomrule
        \end{tabular}
        \caption{Automatic evaluation results for OEQs}\label{tab:res_human_auto}
        \end{subtable}

        \begin{subtable}{0.45\textwidth}
        \centering
        \begin{tabular}{l|c|c}
        \toprule
        & Accuracy & Clarity\\
        \midrule
            Gemini 3 & 61.88 & 69.13 \\
            Gemini 3\textsuperscript{*} & 64.11 & 73.76\\
            GPT-4 & 57.61 & 70.22 \\
            Claude Opus &  \underline{65.89} & \underline{79.43}\\
            \midrule
            Qwen2.5-7B & 57.36 & 67.21 \\
            \bottomrule
        \end{tabular}
        \caption{Human ratings for OEQs}\label{tab:res_human}
        \end{subtable}
        \begin{subtable}{0.45\textwidth}
            \centering
            \begin{tabular}{l|c}   
            \toprule
                Accuracy &  0.54 \\
                Clarity & 0.54\\
                \bottomrule
            \end{tabular}
            \caption{ICC rater agreement (6 classes)}\label{tab:icc}

        \end{subtable}
        
        % \begin{subtable}{0.45\textwidth}
        %     \centering
        %     \begin{tabular}{c|c}
        %     \toprule
        %         Accuracy & Clarity \\
        %         \midrule
        %          66.1 & 96.1 \\
        %          \underline{72.4} & 94.1 \\
        %          71.5 & 95.4 \\
        %          73 & \underline{97.6} \\
        %          \midrule
        %          52.5  & 85.4 \\
        %     \bottomrule
        %     \end{tabular}
        %     \caption{LLM-as-a-judge ratings for OEQs}\label{tab:res_llm}
        % \end{subtable}
\end{table*}
\footnotetext{Under this prompting condition, 5 invalid outputs are identified. Values in parentheses are computed on the valid instances}

%\todo{+ narratives on llma3-70b, qwen3-80b}
On MCQs (see Table \ref{tab:res_mcq1} and \ref{tab:res_mcq2}), the best scores were achieved by Grok-4 with prompts specifying the Hong Kong regulatory context (w/ HK) and Gemini 3 without such specification (w/o HK), and the lowest by DeepSeek-7B under both prompting conditions. The top proprietary LLMs — Gemini 3 and Claude Opus, etc. — have been primarily trained on English corpora. Grok-4’s real-time web searching may have further contributed to its top ranking under the w/ HK context, while simultaneously leading to substantial token consumption. 
In contrast, DeepSeek and Qwen Plus, adapted to Chinese context, were less impressive. This might be explained by the differences in the language distribution of the training data. Our dataset was constructed based on common Hong Kong industry practices, using the English version of the FS Code as a reference. Although both Chinese and English are official languages in Hong Kong, professional examinations (e.g. the PA) are conducted in English, which likely favors models with stronger English training coverage. The results between two prompting conditions aligned with commonsense expectations — specifying the jurisdiction could in general improve accuracy. When jurisdiction was not specified in the prompts, the models might produce longer responses with higher token usage and increased latency, probably in an attempt to reconcile multiple legal systems. 

% 0.5425712726526603 0.5377078352482665
% \begin{table}[htb!]
%     \centering
%      \caption{ICC rater agreement (6 classes).}
%     \label{tab:icc}
%     \begin{tabular}{l|c}   
%     \toprule
%         Accuracy &  0.54 \\
%         Clarity & 0.54\\
%         \bottomrule
%     \end{tabular}
% \end{table}

In contrast, the OEQ results showed a narrower gap between model types across metrics. BLEU scores clustered around 2-3, METEOR and ROUGE mostly fell in the 10-20 range — all relatively low. This is expected, as these metrics emphasize lexical overlap. For OEQs, valid responses may diverge substantially from the reference, leading to lower overlap. Meanwhile, BERTScore values of semantic similarity with the reference were consistently high, which, however, does not help discriminating between models. %\todo{EMM: are the scores already updated? -> lani: scores in table 3 and 4 have been updated; token cost updated in table 1} 

When running the human evaluation, we had to make sure to analyze items for which the annotators were able to reach an agreement, avoiding questions for which even experts gave polarized answers. This led to the exclusion of 32 out of 100, on which the experts showed large disagreements. This filtering step left us with a subset of 68 questions, each one with 5 Q\&A pairs, for a total of 340 pairs. In turn, each of them received two ratings from the experts. 
%\todo{EMM: what is the no. of data points for which we have human ratings again, after filtering? -> lani: now we have 698 pieces of ratings for 68 questions, that is each question having 5 question-answering pairs, and each qa pair gets rated by about 2 times;  the original no. is 1600 ratings for 100 questions} 

The human evaluation revealed narrow accuracy scores around 57 to 65 and clarity scores around 62 to 78. 
For the scores assigned by humans, we computed the intraclass correlation coefficient \citep{koo2016guideline}, obtaining values around 0.54 for both dimensions (see also Table \ref{tab:icc}), which indicates a moderate level of agreement. Considering that both accuracy and clarity are rated on ordinal scales of six values, we deemed it as a sufficiently reliable agreement between our experts.

Regarding the reasoning modes, enabling thinking did not necessarily improve performance. For instance, DeepSeek* (thinking) performed slightly worse than DeepSeek (non-thinking) on OEQs. In other cases, the differences were less pronounced and almost negligible. Yet the resource consumption of the thinking mode was significantly higher: on MCQs, the token usage and response time of DeepSeek* (thinking) were more than five times of DeepSeek (non-thinking). Under the w/ HK context, Gemini 3 (non-thinking) ranked third with an accuracy of 88.5\%, with the lowest cost in terms of both time and token usage. Grok-4, while achieving the highest accuracy, had consumed the largest amount of tokens.

To assess the validity of automatic evaluation metrics, we measured the Spearman correlation between each metric and the human expert scores. Unfortunately, we obtained a negative result: none of the automated metrics showed a significant correlation with expert-rated accuracy, while moderate correlations were observed between clarity and METEOR ($p = 0.38$), ROUGE ($p = 0.36$) and BERTScore ($p = 0.34$).

Compared to industry-standard expectations and requirements, these findings suggest that current general-purpose LLMs cannot yet serve as reliable final recommendations for fire safety design and engineering scenario analysis in the Hong Kong region. Their outputs should only be considered as reference information, subject to supervision and, whenever possible, validation by human experts. It is interesting, in this sense, that experts generally rate LLMs much higher for clarity than for accuracy: in generating an answer, LLMs might be very clear and professional-sounding, but the expressed factual knowledge is not necessarily correct. Moreover, as our correlation analysis shows, traditional automated metrics are not reliable when it comes to the assessment of factual accuracy.

Finally, beyond the observations above, one further implication emerged from our human evaluation. Although there remain many areas for improvement, the evaluation revealed practical challenges. First, interpretations of regulatory provisions can vary — even the official Buildings Department may provide different interpretations — so expert scoring may remain highly subjective despite the provision of references and instructions. Second, experts’ attitudes toward AI outputs differed: some raters, once aware that the text was AI-generated (even without knowing which model), adopted a more skeptical stance, which often resulted in lower scores. Future initiatives — whether academic projects or industry application — should account for the inherent subjectivity of human ex- perts and their attitudes toward AI outputs.

 % # True 0.4045048771827044 0.5643752715036062  # 等分
 %    #  True 0.3332320046908876 0.5562178470498959  # 不等分
 %    # False 0.46268007922575843 0.6205671431120897 # 不分

\section{Conclusion}

This work presents the first systematic evaluation of 12 LLMs, both proprietary and open-source, on their ability to interpret the Hong Kong fire safety code. To support this study, we collaborated with fire safety experts in Hong Kong to construct a Q\&A dataset comprising both MCQs and OEQs, emulating the structure of official professional examination.

Overall, proprietary models outperform  open-source ones. On MCQs, proprietary models  achieved about 80–90\% accuracy on MCQs,  while open-source models performed substantially  lower. On OEQs, model performance remains  limited. Ratings from human experts %and the LLM  judge 
 on OEQs reveal that accuracy of model  outputs remain only around 50–70\%, with most  responses falling into the major issues category  rather than reaching fully accurate or minor  inaccuracy levels. This indicates that current  LLMs cannot yet reliably capture the precision  required for regulatory compliance. It is also found  that enabling thinking did not yield significant  gains on either setting, but incurred high resource  consumption, underscoring the need to balance  efficiency with effectiveness. Moreover, enabling  thinking modes did not yield significant accuracy  gains in either MCQs or OEQs, but increased  resource consumption. Finally, all automated metrics were not correlated with accuracy and only moderately correlated with clarity, thus confirming the challenges in evaluation when the feedback of human experts is not easily available.

While constrained by dataset scarcity and limited  expert availability, this study contributes a localized  benchmark for evaluating LLMs against the Hong  Kong fire safety code. 

Although our focus is on fire  safety, the underlying methodology is transferable to other AEC subdomains. For future work, we will prioritize  improving evaluation frameworks — especially human evaluation, and expand the benchmark beyond  fire safety to include broader regulatory tasks in  AEC. 

\section*{Limitations}
We acknowledged the limitations of the current work are primarily: First, since fire safety experts are difficult to recruit and our dataset requires manual review, screening and revision to ensure quality and accuracy, we are aware of the fact that the dataset size is small. This issue, together with the specific focus on the Hong Kong regulation, is an important limitation to the the generalizability of the findings. The same challenge of expert recruitment also limited the coverage of human evaluation.%, for which we mostly had to rely on LLM as judges.

\section*{Ethics Statement}
\textbf{Potential risk}: The potential risk of this work lies in misinterpretation of compliance-checking outputs, which could lead to incorrect conclusions about safety compliance. To mitigate this, we emphasize that LLM predictions are advisory only and must be validated by certified fire safety professionals. 

\textbf{License}: All artifacts (LLMs,  data, and software packages) were used in accordance with their intended use and licensing terms in our dataset construction and evaluation experiments. Proprietary models (Claude Opus, Qwen Plus, GPT‑5, GPT‑4, Gemini 3, DeepSeek) were accessed via licensed APIs in compliance with providers’ terms of service and are not freely re-distributable. Open‑source models were used under their respective permissive licenses (e.g., Apache 2.0, MIT), ensuring reproducibility. Other open-source tools (such as \texttt{pycocoevalcap}) are released under permissive licenses. 

\textbf{Data Safety}: Our dataset is derived from publicly available regulatory documents and has been manually reviewed to ensure that it does not contain personally identifying information or offensive content. Human evaluation data was also collected under voluntary participation. Personal identifying information such as names is removed to protect privacy. 

\textbf{Expert participation}: Human experts were provided with clear instructions outlining the evaluation task, including definitions, examples, and scoring rubrics (an example survey is presented in Appendix \ref{sec:survey_tmp}). They were informed of the study purpose and how the data would be collected and used. The participation was voluntary and all participants retained their rights to withdraw or remove their data at any point. As compensation, domain‑specific workshops were offered. All participants gave informed consent before contributing.

\textbf{AI Assistance}: Copilot was used to support writing in language polishing.
\section*{Acknowledgment}

EC acknowledges support from the project “Analyzing the semantics of Transformers representations for financial natural language processing” (ZVYU), funded by the Faculty of Humanities of the Hong Kong Polytechnic University
(PolyU-UGC). We would also like to thank our
reviewers for the constructive feedback.

Special thanks are also due to all contributors who assisted with dataset creation and evaluation.

\par\textbf{Dataset Creation}:
\begin{itemize}
    \item Ar. Chau Lok Kan, Director, RA(HK) HKIA — Co-creator
    \item Ir. Zhao Yueci, Associate, BEng CPEng NPER MIEAust MHKIE — Contributor
\end{itemize}
\par\textbf{Human Evaluation} (all as Contributors):
\begin{itemize}
    \item Ar Dr. Adrian Lo, RA(HK) HKIA RIBA
    \item Ar. Amy Chan Nga Yin, Senior Project Manager, RA(HK) HKIA Authorized Person
    \item Ar. Emilie Wong, Founder, RA(HK) HKIA RIBA EDAC
    \item Ar. Jessica Chong, Assistant Manager, RA(HK) HKIA
    \item Ar. Jessica Kong, Director, RA(HK) HKIA
    \item Ar. Jiang Hejia, Co-founder, RA(HK) HKIA
    \item Ar. Lily Yu, Design Manager, RA(HK) HKIA
    \item Ar. Lu Ruixing, Senior Associate, RA(PRC)
    \item Ar. Nam Wu, RA(HK) HKIA
    \item Ir. Pin Wai Kei, Technical Director, MHKIE MIFireE
    \item Ar. Simon Chan, Lecturer, RA(HK) HKIA
    \item Ar. Song Huiqing, Architect, RA(HK) HKIA
    \item Ar. Sun Bochao, Architect, RA(HK) HKIA
    \item Ir. Sunny Wong Ho Ching, MHKIE RPE(Fire) CEng MIMEchE MCIBSE BEAM Pro CESGA
    \item Ar. Tian Yi, Senior Associate, RA(PRC) RA(HK) HKIA
    \item Ir. Victor Tung Kok Keung, Retired Fire Officer, RPE(Fire)
    \item Ar. Vincent Choi Kei Shun, Design Director, RA(HK) HKIA
    \item Ar. Xia Meng Yuan, Senior Associate, RA(HK) HKIA
    \item Ar. Xu Wei, Associate, RA(HK) HKIA
    \item Ar. Zeng Jingyao, Architect, RA(HK) HKIA
    \item Ar. Zhang Shuang, Architect, RA(HK) HKIA
\end{itemize}

%\section{References}
\bibliography{ref}

\begin{thebibliography}{27}
\providecommand{\natexlab}[1]{#1}

\bibitem[{Anthropic(2024)}]{anthropic2024opus}
Anthropic. 2024.
\newblock Claude opus.
\newblock \url{https://www.anthropic.com}.
\newblock Large language model.

\bibitem[{Banerjee and Lavie(2005)}]{banerjee-lavie-2005-meteor}
Satanjeev Banerjee and Alon Lavie. 2005.
\newblock \href {https://aclanthology.org/W05-0909/} {{METEOR}: An automatic metric for {MT} evaluation with improved correlation with human judgments}.
\newblock In \emph{Proceedings of the {ACL} Workshop on Intrinsic and Extrinsic Evaluation Measures for Machine Translation and/or Summarization}, pages 65--72, Ann Arbor, Michigan. Association for Computational Linguistics.

\bibitem[{Chae et~al.(2026)Chae, Yeom, Park, Bae, Jang, Jin, Jang, and Kim}]{chae2026evaluating}
Kyubyung Chae, Jewon Yeom, Jeongjae Park, Seunghyun Bae, Ijun Jang, Hyunbin Jin, Jinkwan Jang, and Taesup Kim. 2026.
\newblock {Evaluating Structure-Aware Retrieval and Safety in Statute-Centric Legal QA}.
\newblock In \emph{Proceedings of ACL}.

\bibitem[{DeepMind(2025)}]{gemini32025}
Google DeepMind. 2025.
\newblock Gemini 3.
\newblock \url{https://deepmind.google/technologies/gemini}.
\newblock Large language model.

\bibitem[{{DeepSeek-AI}(2024)}]{deepseek2024}
{DeepSeek-AI}. 2024.
\newblock Deepseek-v3 technical report.

\bibitem[{Del~Campo and Leach(2022)}]{del2022machine}
Matias Del~Campo and Neil Leach. 2022.
\newblock \emph{{Machine Hallucinations: Architecture and Artificial Intelligence}}.
\newblock John Wiley \& Sons.

\bibitem[{Ghimire et~al.(2024)Ghimire, Kim, and Acharya}]{buildings14010220}
Prashnna Ghimire, Kyungki Kim, and Manoj Acharya. 2024.
\newblock {Opportunities and Challenges of Generative AI in Construction Industry: Focusing on Adoption of Text-Based Models}.
\newblock \emph{Buildings}, 14(1).

\bibitem[{Guo et~al.(2025)Guo, Li, Kong, Jiang, Zhao, Gong, Zhang, Li, Sang, Zhu et~al.}]{guo2025toward}
Xingang Guo, Yaxin Li, Xiangyi Kong, Yilan Jiang, Xiayu Zhao, Zhihua Gong, Yufan Zhang, Daixuan Li, Tianle Sang, Beixiao Zhu, and 1 others. 2025.
\newblock {Toward Engineering AGI: Benchmarking the Engineering Design Capabilities of LLMs}.
\newblock \emph{arXiv preprint arXiv:2509.16204}.

\bibitem[{Hettiarachchi et~al.(2026)Hettiarachchi, Gaber, Parsafard, and Vakaj}]{hettiarachchi-etal-2026-snowtec}
Hansi Hettiarachchi, Mohamed~Medhat Gaber, Pouyan Parsafard, and Edlira Vakaj. 2026.
\newblock {SNOWTEC: Synthetic Natural language Oversampling With Transformer-based Information ExtraCtion for Automated Compliance Checking}.
\newblock \emph{Machine Learning with Applications}, 24:100911.

\bibitem[{Huang et~al.(2023)Huang, Yu, Ma, Zhong, Feng, Wang, Chen, Peng, Feng, Qin et~al.}]{huang2023survey}
Lei Huang, Weijiang Yu, Weitao Ma, Weihong Zhong, Zhangyin Feng, Haotian Wang, Qianglong Chen, Weihua Peng, Xiaocheng Feng, Bing Qin, and 1 others. 2023.
\newblock {A Survey on Hallucination in Large Language Models: Principles, Taxonomy, Challenges, and Open Questions}.
\newblock \emph{ACM Transactions on Information Systems}.

\bibitem[{Hurst et~al.(2024)Hurst, Lerer, Goucher, Perelman, Ramesh, Clark, Ostrow, Welihinda, Hayes, Radford et~al.}]{hurst2024gpt}
Aaron Hurst, Adam Lerer, Adam~P Goucher, Adam Perelman, Aditya Ramesh, Aidan Clark, AJ~Ostrow, Akila Welihinda, Alan Hayes, Alec Radford, and 1 others. 2024.
\newblock Gpt-4o system card.
\newblock \emph{arXiv preprint arXiv:2410.21276}.

\bibitem[{Kampelopoulos et~al.(2025)Kampelopoulos, Tsanousa, Vrochidis, and Kompatsiaris}]{kampelopoulos2025review}
Dimitrios Kampelopoulos, Athina Tsanousa, Stefanos Vrochidis, and Ioannis Kompatsiaris. 2025.
\newblock {A Review of LLMs and their Applications in the Architecture, Engineering and Construction Industry}.
\newblock \emph{Artificial Intelligence Review}, 58(8):250.

\bibitem[{Koo and Li(2016)}]{koo2016guideline}
Terry~K Koo and Mae~Y Li. 2016.
\newblock A guideline of selecting and reporting intraclass correlation coefficients for reliability research.
\newblock \emph{Journal of Chiropractic Medicine}, 15(2):155--163.

\bibitem[{Liang et~al.(2026)Liang, Huang, Wang, Chai, Yu, Wei, Liu, Li, Wang, Luo et~al.}]{liang2026aecbench}
Chen Liang, Zhaoqi Huang, Haofen Wang, Fu~Chai, Chunying Yu, Huanhuan Wei, Zhengjie Liu, Yanpeng Li, Hongjun Wang, Ruifeng Luo, and 1 others. 2026.
\newblock {AECBench: A Hierarchical Benchmark for Knowledge Evaluation of Large Language Models in the AEC Field}.
\newblock \emph{Advanced Engineering Informatics}, 71:104314.

\bibitem[{Lin(2004)}]{lin-2004-rouge}
Chin-Yew Lin. 2004.
\newblock \href {https://aclanthology.org/W04-1013/} {{ROUGE}: A package for automatic evaluation of summaries}.
\newblock In \emph{Text Summarization Branches Out}, pages 74--81, Barcelona, Spain. Association for Computational Linguistics.

\bibitem[{Papineni et~al.(2002)Papineni, Roukos, Ward, and Zhu}]{papineni-etal-2002-bleu}
Kishore Papineni, Salim Roukos, Todd Ward, and Wei-Jing Zhu. 2002.
\newblock \href {https://doi.org/10.3115/1073083.1073135} {{B}leu: a method for automatic evaluation of machine translation}.
\newblock In \emph{Proceedings of the 40th Annual Meeting of the Association for Computational Linguistics}, pages 311--318, Philadelphia, Pennsylvania, USA. Association for Computational Linguistics.

\bibitem[{{Qwen Team}(2024)}]{qwen2.5}
{Qwen Team}. 2024.
\newblock \href {https://qwenlm.github.io/blog/qwen2.5/} {Qwen2.5: A party of foundation models}.

\bibitem[{{Qwen Team}(2025)}]{qwen3technicalreport}
{Qwen Team}. 2025.
\newblock \href {https://arxiv.org/abs/2505.09388} {Qwen3 technical report}.
\newblock \emph{Preprint}, arXiv:2505.09388.

\bibitem[{{Qwen Team}(2026)}]{qwen36plus}
{Qwen Team}. 2026.
\newblock \href {https://qwen.ai/blog?id=qwen3.6} {{Qwen3.6-Plus}: Towards real world agents}.

\bibitem[{Singh et~al.(2025)Singh, Fry, Perelman, Tart, Ganesh, El-Kishky, McLaughlin, Low, Ostrow, Ananthram et~al.}]{singh2025openai}
Aaditya Singh, Adam Fry, Adam Perelman, Adam Tart, Adi Ganesh, Ahmed El-Kishky, Aidan McLaughlin, Aiden Low, AJ~Ostrow, Akhila Ananthram, and 1 others. 2025.
\newblock {OpenAI GPT-5 System Card}.
\newblock \emph{arXiv preprint arXiv:2601.03267}.

\bibitem[{xAI(2024)}]{xai2024grok4}
xAI. 2024.
\newblock Grok-4.
\newblock \url{https://x.ai}.
\newblock Large language model.

\bibitem[{Xiong et~al.(2026)Xiong, Netser, Tang, Li, and Hwang}]{xiong2026socio}
Ruoxin Xiong, Yael Netser, Pingbo Tang, Beibei Li, and Joonsun Hwang. 2026.
\newblock {Socio-technical Assessment of Generative AI Integration in Architecture, Engineering, and Construction (AEC) Workflows: An Empirical Study Using O* NET Occupational Taxonomy}.
\newblock \emph{Advanced Engineering Informatics}, 71:104392.

\bibitem[{Yang et~al.(2024)Yang, Yang, Hui, Zheng, Yu, Zhou, Li, Li, Liu, Huang, Dong, Wei, Lin, Tang, Wang, Yang, Tu, Zhang, Ma, Xu, Zhou, Bai, He, Lin, Dang, Lu, Chen, Yang, Li, Xue, Ni, Zhang, Wang, Peng, Men, Gao, Lin, Wang, Bai, Tan, Zhu, Li, Liu, Ge, Deng, Zhou, Ren, Zhang, Wei, Ren, Fan, Yao, Zhang, Wan, Chu, Liu, Cui, Zhang, and Fan}]{qwen2}
An~Yang, Baosong Yang, Binyuan Hui, Bo~Zheng, Bowen Yu, Chang Zhou, Chengpeng Li, Chengyuan Li, Dayiheng Liu, Fei Huang, Guanting Dong, Haoran Wei, Huan Lin, Jialong Tang, Jialin Wang, Jian Yang, Jianhong Tu, Jianwei Zhang, Jianxin Ma, and 40 others. 2024.
\newblock Qwen2 technical report.
\newblock \emph{arXiv preprint arXiv:2407.10671}.

\bibitem[{Yang et~al.(2025)Yang, Yu, Li, Liu, Huang, Huang, Jiang, Tu, Zhang, Zhou, Lin, Dang, Yang, Yu, Li, Sun, Zhu, Men, He, Xu, Yin, Yu, Qiu, Ren, Yang, Li, Xu, and Zhang}]{qwen2.5-1m}
An~Yang, Bowen Yu, Chengyuan Li, Dayiheng Liu, Fei Huang, Haoyan Huang, Jiandong Jiang, Jianhong Tu, Jianwei Zhang, Jingren Zhou, Junyang Lin, Kai Dang, Kexin Yang, Le~Yu, Mei Li, Minmin Sun, Qin Zhu, Rui Men, Tao He, and 9 others. 2025.
\newblock Qwen2.5-1m technical report.
\newblock \emph{arXiv preprint arXiv:2501.15383}.

\bibitem[{Zhang et~al.(2020)Zhang, Kishore, Wu, Weinberger, and Artzi}]{zhang2020bertscore}
Tianyi Zhang, Varsha Kishore, Felix Wu, Kilian~Q. Weinberger, and Yoav Artzi. 2020.
\newblock Bertscore: Evaluating text generation with bert.
\newblock In \emph{International Conference on Learning Representations}.

\bibitem[{Zheng et~al.(2025)Zheng, Wong, Su, Tang, Nawaz, and Kassem}]{zheng2025automating}
Chunmo Zheng, Saika Wong, Xing Su, Yinqiu Tang, Ahsan Nawaz, and Mohamad Kassem. 2025.
\newblock {Automating Construction Contract Review Using Knowledge Graph-enhanced Large Language Models}.
\newblock \emph{Automation in Construction}, 175:106179.

\bibitem[{Zheng et~al.(2023)Zheng, Chen, Cao, Lu, and Lin}]{zheng2023llm}
Zhe Zheng, Ke-Yin Chen, Xin-Yu Cao, Xin-Zheng Lu, and Jia-Rui Lin. 2023.
\newblock {LLM-funcmapper: Function Identification for Interpreting Complex Clauses in Building Codes via LLM}.
\newblock \emph{arXiv preprint arXiv:2308.08728}.

\end{thebibliography}

\appendix
\label{sec:appendix}
\onecolumn
\newpage
\section{Supplementary Tables}

\begin{longtable}{lp{0.75\linewidth}}
    \caption{Example instances from the dataset.} \label{tab:data_exp}\\
    \toprule
    \endfirsthead

    \multicolumn{2}{r}{\textit{(Table continues from previous page)}}\\
    \midrule
    \endhead

    \midrule
    \multicolumn{2}{r}{\textit{(Table continues on the next page)}}\\
    \endfoot

    \bottomrule\\
    \endlastfoot

    Question: & What is the main purpose of a Qualitative Design Review (QDR) in a performance-based fire safety design?\\
    & A) To confirm building materials meet combustibility limits\\
    & B) To set fire safety objectives, identify design scenarios, choose acceptance criteria, and outline analysis methods\\
    & C) To run computational fluid dynamics (CFD) simulations\\
    & D) To commission fire service installations\\
    
    Reference Answer:  & B\\
    Explanation: & A qualitative design review establishes the fire safety objectives, scenarios, acceptance criteria, and the analysis approach before any detailed quantitative work is carried out.\\
    % Difficulty Level: & 1 \\
    \midrule
    Question: & An office tenant proposes to shorten the escape path by routing occupants through a storage room and staff pantry to reach the exit stair. The storage is used for paper and small solvent supplies and is kept locked during normal hours. Which of the following design options best meets the means of escape requirement?\\
    & A) Keep the route but add illuminated exit signs and emergency lighting through the storage to replace the visual fire alarm inside the office\\
    & B) Reroute the escape path so it avoids rooms with higher fire load and provide a protected corridor where necessary to connect to the exit stair\\
    & C) Allow the route but restrict it to staff via access control during evacuation\\
    & D) Keep the route but install a wider door from the storage to the stair, keep the door from the office to the storage opaque, additional visual fire alarms not required \\
    Reference Answer: & B\\
    Explanation: & Escape routes should not pass through rooms with elevated fire risk (e.g., storerooms, kitchens/pantries). The intent is that occupants move within protected routes separated from fire hazards until reaching a place of safety. Signage, door width, or administrative controls cannot compensate for the inherent hazard of routing through high-risk rooms; a protected corridor or lobby is used to maintain separation where needed.\\
    % Difficulty Level: & 3 \\
    % Engineering-Related & \\
    \midrule
    Question: & A restaurant with an estimated 80 patrons uses a frameless glass pivot door on its main exit that swings inward into the dining area to avoid obstructing the sidewalk. Is this acceptable, and what modification is required? \\
    Reference Answer: & This is not acceptable because doors serving a floor or storey for more than 30 persons must swing in the direction of egress. In case the doors are swinging in both ways upper view panels shall be provided. Inward-swinging doors impede escape under crowd pressure and can trap occupants and are not allowed for rooms or storeys accommodating more than 30 persons. The door should be re-hung to swing outward (toward the place of safety) and be readily openable without a key or special knowledge. If sidewalk clearance is a concern, provide a small vestibule with an outward-swinging exit door while maintaining the required clear exit width. \\
    Explanation: & The intent is to prevent crowd surge from blocking egress and to ensure doors can be opened quickly during an emergency. The outward swing requirement for exit doors is triggered by higher occupant loads, and doors must be openable without delay or special action. Any redesign must preserve the minimum clear exit width and avoid hardware or stops that reduce it.\\
    % Difficulty Level: & 2 \\
    Engineering-Related: & No \\
    \midrule
    Question: & A performance-based smoke control design is proposed for a tall atrium to reduce the smoke exhaust rates. What must be demonstrated for the design to be acceptable, and how is occupant safety evaluated?\\
    Reference Answer: & Define credible design fire scenarios and use validated models to predict smoke production, movement, and layer heights. Show that the available safe egress time exceeds the required safe egress time by an appropriate margin, considering pre-movement and evacuation durations. Demonstrate tenability along evacuation paths (visibility, temperature, and toxicity) remains within accepted limits at head height for the evacuation period. Provide sensitivity analyses for key parameters, document system reliability and management provisions, and demonstrate equivalence to prescriptive intent.\\
    Explanation: & Performance-based acceptance hinges on meeting fire safety objectives through comparison of ASET and RSET and confirming tenability criteria under plausible worst cases. Validated modeling and sensitivity analyses ensure robustness against uncertainty. Management and reliability measures uphold the performance assumptions throughout the building’s life.\\
    % Difficulty Level: & 3 \\
    Engineering-Related: & Yes \\
\end{longtable}
\clearpage

\begin{table}[htb!]
    \centering
    \caption{Templates of Evaluation Prompts}\label{tab:prompts_eval}
        \begin{subtable}{\textwidth}
        \centering
        \begin{tabular}{p{0.9\linewidth}}
        \toprule
             \texttt{You are answering a question about the Hong Kong fire safety code (Code of Practice for Fire Safety in Buildings 2011).} \colorbox{gray!30}{[PLACEHOLDER]} \texttt{Return the result strictly as a JSON object in the following format: \{"answer": "xx", "reason": "xx"\}}\quad\colorbox{gray!30}{[QUESTION]}\\
        \bottomrule
    \end{tabular}
    \caption{Template for OEQs and MCQs with Hong Kong context specified\protect\footnotemark}\label{tab:tmp_oeq}
    \end{subtable}
    \begin{subtable}{\textwidth}
        \centering
        \begin{tabular}{p{0.9\linewidth}}
        \toprule
             \texttt{You are given a multiple‑choice question. Choose ONE most appropriate answer from the options. Return the result strictly as a JSON object in the following format: \{"answer": "xx", "reason": "xx"\}}\quad\colorbox{gray!30}{[QUESTION]}\\
        \bottomrule
    \end{tabular}
    \caption{Template for MCQs without Hong Kong context (w/o HK)}\label{tab:tmp_mq2}
    \end{subtable}
\end{table}
\footnotetext{The \colorbox{gray!30}{[PLACEHOLDER]} is determined by the question type. For OEQs, it is filled with \texttt{You are given a short-answer question. Respond in a concise, accurate, and professional manner.} Otherwise, it is \texttt{You are given a multiple‑choice question. Choose ONE most appropriate answer from the options. }}

\section{Survey Template for Human Evaluation}
\label{sec:survey_tmp}
For demonstration purposes, we created a simple survey template using Google Form. Any sections that could potentially reveal author information have been concealed. All expert raters (architects and fire engineers) were recruited, contacted, and invited to participate by two of the authors. Information such as evaluation purpose and instructions were personally communicated to participants (either face-to-face, via email, or through other means of contact). To protect participant privacy, identifying information has been hidden.

\includepdf[pages=-]{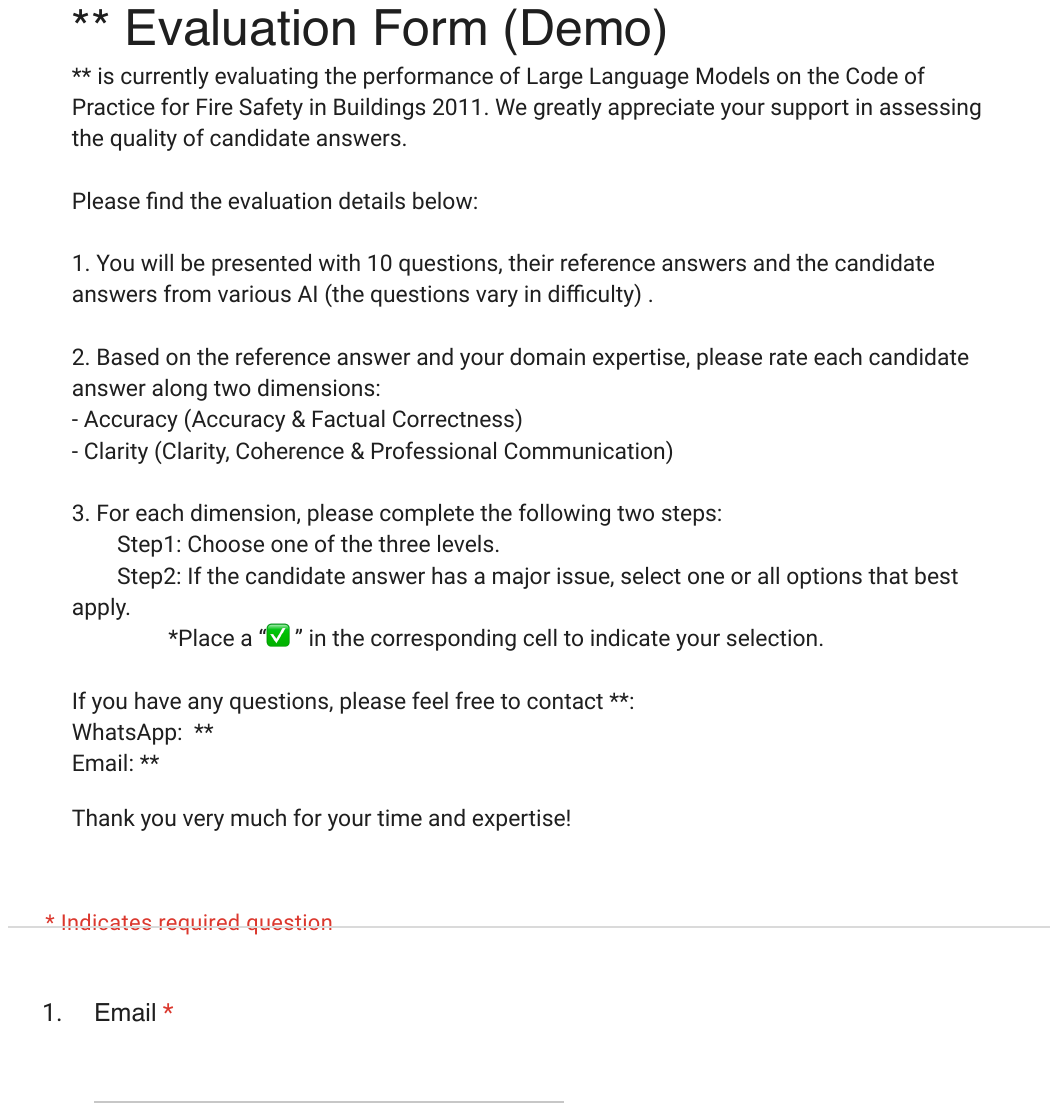}

\end{document}